\documentclass[pre,10pt,amsmath,amssymb]{revtex4}
\usepackage{graphicx}

\usepackage{amsmath,amsfonts,amssymb,bm}

\begin{document}

\title{Bistable generalised Langevin dynamics driven by correlated noise possessing a long jump distribution: 
 barrier crossing and stochastic resonance}

\author
{Tomasz Srokowski}

\affiliation{
 Institute of Nuclear Physics, Polish Academy of Sciences, PL -- 31-342
Krak\'ow,
Poland }


\begin{abstract} 

The generalised Langevin equation with a retarded friction and 
a double-well potential is solved. The random force is modelled by 
a multiplicative noise with long jumps. 
Probability density distributions converge with time to 
a distribution similar to a Gaussian but tails have a power-law form. Dependence 
of the mean first passage time on model parameters is discussed. Properties 
of the stochastic resonance, emerging as a peak in the plot of the spectral amplification 
against the temperature, are discussed for various sets of the model parameters. 
The amplification rises with the memory and is largest for the cases corresponding to 
the large passage time. 

\end{abstract} 


\maketitle

\section{Introduction}

The stochastic dynamics may not always be restricted to familiar problems involving 
Gaussian distributions. In realistic systems long jumps are frequently observed 
and the variance may be infinite, as it is the case for the general L\'evy stable 
distributions which possess power-law tails. Since a medium nonhomogeneity is 
typical for systems exhibiting long jumps, one can expect that a random force 
in a dynamical description of those systems depends on the process value, i.e. 
the stochastic equation contains a multiplicative noise. An interesting property 
of such description is a possibility that the variance acquires a finite value even if 
the underlying process is assumed as a non-Gaussian L\'evy stable process, 
defined by the stability index $\alpha$ $(0<\alpha<2)$ \cite{sro09}. 

The covariance functions are finite as well. The autocorrelation function for 
the generalised Ornstein- Uhlenbeck process $\xi(t)$, defined by the Langevin 
equation driven by a multiplicative Cauchy 
noise, falls with time like a stretched exponential but may be reasonable 
approximated by a simple exponential \cite{sron}. Then the process $\xi(t)$ comprises 
two features: it exhibits long jumps and possesses well-defined covariance functions. 
When such a process enters a stochastic equation as a random driving, 
the finite correlation time requires a usual damping term to be 
substituted by a retarded friction to ensure a proper equilibrium state. This 
problem is well-known for the Gaussian processes: one can construct an effective low-dimensional 
equation describing a many-body system in which individual variables are coupled by 
harmonic oscillators \cite{mori}. The effective equation -- the generalised Langevin 
equation (GLE) -- is non-local in time and satisfies the fluctuation-dissipation theorem 
\cite{kubo}. Moreover, the effective random force is Gaussially distributed. 
If velocities of bath particles are non-Gaussian but of finite variance, 
one may still expect convergence to the Gaussian, according to the central limit 
theorem. However, this theorem does not apply in the presence of long range correlations, 
which is natural in complex systems. On the other hand, large higher 
moments make convergence to the Gaussian impossible even if independent variables are linearly combined. 
In particular, the distance of that distribution from the Gaussian is infinite when 
the third moment diverges, according to the Berry-Ess\'een theorem \cite{fel}. 
Therefore, the effective noise may be non-Gaussian and the dynamics of such systems may still 
be described  by GLE  \cite{cof}. GLE driven by $\xi(t)$ for the case 
without a deterministic force was analysed in Ref.\cite{sron}. The probability density 
distributions converge with time to a stationary state with a power-law tail; 
the variance grows linearly with time indicating a normal diffusion. 

In this paper, we consider GLE driven by $\xi(t)$ for 
the case of a bistable system and activated, in addition, 
by a periodic force. We address a physically important problem of the time 
characteristics of the barrier penetration and analyse the influence of 
the memory on this process. We calculate, in particular, the mean 
first passage time (MFPT) as a function of model parameters. When the rate of the jumping 
between the potential wells due to the noise coincides with the frequency of the 
oscillatory force, the stochastic resonance (SR) is observed. We discuss this phenomenon 
and demonstrate how model parameters modify its properties, in particular 
the position and intensity.

\section{Stochastic equations and density distributions}

The ordinary Ornstein-Uhlenbeck process describes motion of the particle subjected to 
a linear deterministic force and an additive Gaussian white noise. It can be generalised 
by admitting a dependence of the noise on the process value (a multiplicative noise) 
and including non-Gaussian distributions, in particular the L\'evy stable 
distributions with power-law tails. Then the generalised process obeys the following 
Langevin equation 
\begin{equation}
\label{la}
\dot\xi(t)=-\gamma\xi(t)+G(\xi)L(t), 
\end{equation}
where $L(t)$ is the noise and a given function $G(\xi)$ may be responsible 
e.g. for a nonhomogeneous structure of the environment. Since $L(t)$ is a white noise, 
Eq.(\ref{la}) is not unique and requires a clarification at which time $G(\xi)$ is to be evaluated; in the following 
the Stratonovich interpretation will be applied. We assume 
\begin{equation}
\label{godxi}
G(\xi)=K|\xi|^{-\theta}, 
\end{equation}
where the constant $K$ has the dimension $[K]=$cm$^\theta$. The algebraic form of Eq.(\ref{godxi}) 
is well suited to describe e.g. self-similar systems. Moreover, distribution of the independent increments of $L(t)$ 
is assumed in a form of the symmetric Cauchy distribution, 
\begin{equation}
\label{cau}
p_C(x)=\frac{1}{\pi}\frac{D}{D^2+x^2}, 
\end{equation}
where a generalised diffusion coefficient $D$ has the dimension $[D]=$ cm/sec. All moments 
of the distribution (\ref{cau}) are infinite. In the Stratonovich interpretation, a distribution 
corresponding to Eq.(\ref{la}) can be easily derived by introducing a new variable, 
\begin{equation}
\label{yodx}
\eta=\frac{K^{-1}}{1+\theta}|\xi|^{1+\theta}\hbox {sgn}(\xi),  
\end{equation}
which transforms Eq.(\ref{la}) to an equation with the additive noise, 
\begin{equation}
\label{laa}
\dot\eta(t)=-\gamma_\theta\eta(t)+L(t), 
\end{equation}
where $\gamma_\theta=\gamma(1+\theta)$ \cite{sro09}. Distribution of the original variable reads 
\begin{equation}
\label{conprc}
p_\xi(\xi,t)=\frac{|\xi|^\theta}{DK\pi\gamma_\theta}\frac{1-\exp(-\gamma_\theta t)}
{\xi^{2+2\theta}/D^2K^2(1+\theta^2)+(1-\exp(-\gamma_\theta t))^2/\gamma_\theta^2};
\end{equation}
the initial condition is $p_\xi(\xi,0)=\delta(\xi)$. 
The variance exists if $\theta>1$ and it can be strictly derived \cite{sro09}. 
The formula for the autocorrelation function, in turn, requires an integration of the process values 
with the transition probability, 
\begin{equation}
\label{defc}
{\cal C}(t)=\langle\xi(0)\xi(t)\rangle=
\int\int\xi_1\xi_2p_\xi(\xi_2,t;\xi_1,0)d\xi_1d\xi_2=\int\int\xi_1\xi_2p_\xi(\xi_2,t|\xi_1,0)p_\xi(\xi_1)d\xi_1d\xi_2,
\end{equation}
where $p_\xi(\xi)=\lim_{t\to\infty}p_\xi(\xi,t)$ and $p_\xi(\xi_2,t|\xi_1,0)$ is the conditional probability. 
One can prove that the asymptotic time-dependence of ${\cal C}(t)$ is exponential with the rate $\gamma$ but only 
if one introduces a truncation of the distribution (\ref{cau}). Otherwise, a numerical evaluation of 
the double integral (\ref{defc}) reveals a stretched exponential shape at large time \cite{sron}. However, 
a deviation of ${\cal C}(t)$ from the simple exponential is very small and the dependence of the rate on 
both system parameters, $\gamma$ and $\theta$, appears simple. Therefore, we may approximate the covariance 
by the following expression, 
\begin{equation}
\label{cex}
{\cal C}(t)=(KD)^{2/(1+\theta)}\frac{\gamma^{-2/(1+\theta)}}{\cos(\pi/(1+\theta))}\hbox{e}^{-\lambda t}, 
\end{equation}
where $\lambda=0.80(\theta+0.31)\gamma$ follows from the numerical analysis. 

We consider a dynamical system described by GLE which is driven by the process $\xi(t)$ and 
contains a deterministic force $F(x,t)=-\partial V(x,t)/\partial x$. Therefore, $\xi$ is now 
interpreted as a force with the dimension $[\xi]=\hbox{g cm/sec}^2$ and, accordingly, 
$[K]=(\hbox{g cm/sec}^2)^\theta$ and $[D]=\hbox{g cm/sec}^3$. 
The equation is the following: 
\begin{equation}
\label{gle}
m \frac{dv(t)}{dt} = F(x,t)-m\int_0^t K(t-\tau)v(\tau)d\tau + \xi(t), 
\end{equation}
where $v(t)$ is a velocity and $m$ denotes the particle mass. The memory kernel is related to 
the noise autocorrelation function by the second fluctuation-dissipation theorem, 
$K(t)={\cal C}(t)/mT$, where $T$ is the temperature and the Boltzmann constant is set at one. 
The correlation time parameter $\lambda$ is determined not only by the damping $\gamma$, 
as it is the case for the ordinary Ornstein-Uhlenbeck process, but also by $\theta$. 
Since $1/\lambda$ is to be interpreted as the memory time, we require that ${\cal C}(t)$ 
converges to the delta function in the limit $\lambda\to\infty$. Therefore, we rescale 
the driving noise: $\xi(t)\rightarrow c_1\sqrt{\lambda}\xi(t)$, 
where $c_1=\sqrt{0.5411{\cal C}(0)}$ has the dimension $\sqrt{\hbox{sec}}$, and appropriately modify the kernel. 
Mathematically, Eq.(\ref{gle}) represents an integro-differential equation of the Volterra type 
which is nonlinear in general; it cannot be solved by an integral transforms technique and we must 
resort to numerical methods. GLE is easier to handle when we substitute Eq.(\ref{gle}) 
by a system of the differential equations which a procedure is possible for the exponential 
$C(t)$. A simple derivation yields 
\begin{eqnarray}
\label{gled}
m\dot v(t)&=&F(x,t)-mw(t)+c_1\sqrt{\lambda}\xi(t)\nonumber\\
\dot w(t)&=&-\lambda w(t)+ c_1^2{\cal C}(0)\lambda v(t)/mT \nonumber\\
\dot x(t)&=&v(t)\nonumber\\
\dot \eta(t)&=&-\gamma(1+\theta)\eta(t)+L(t). 
\end{eqnarray}
We assume the initial conditions: $x(0)=x_0$, $v(0)=v_0$, $w(0)=0$ and $\eta(0)$ 
is given by a stationary distribution for the process (\ref{laa}). Moreover, we assume 
in the numerical calculations $D=1\, \hbox{g cm/sec}^3$ and $K=1\, (\hbox{g cm/sec}^2)^\theta$. 
GLE in a form similar to (\ref{gled}) is frequently studied for the Gaussian processes, 
e.g. for such problems as the transport of the Brownian 
particles \cite{lucz1,lucz}, non-Markovian features of the nonlinear GLE \cite{far} and 
a fractional superdiffusion \cite{sie}. 
In those cases the system (\ref{gled}) represents an equivalent 
Markovian dynamics in a higher-dimensional space which may be expressed by a multidimensional 
Fokker-Planck equation. The present case is more complicated, even in the absence of the 
potential, because of the non-linear relation between $\eta$ and $\xi$, Eq.(\ref{yodx}). 

First, let us consider a time-independent bistable potential, 
\begin{equation}
\label{pot}
V(x)=-\frac{A}{2}x^2+\frac{B}{4}x^4,
\end{equation}
which has two minima positioned at $x_m=\pm\sqrt{A/B}$. In the numerical 
calculations, we assume $A=4\, \hbox{g/sec}^2$ and $B=1\, \hbox{g/cm}^2\hbox{sec}^2$. 
The distributions were derived by a numerical solving of the system (\ref{gled}) and the integration 
was performed by applying the Heun method. The time evolution of the distributions 
is presented in Fig.1. The velocity distribution, initially 
centred at the origin, widens with time to develop two peaks at large $|v|$. Finally, it converges 
to a stationary state which assumes an apparent Maxwellian shape. However, a magnification 
of the tail indicates a clear power-law asymptotics, similarly to the case without 
the potential \cite{sron}. The stationary position distribution exhibits two peaks 
corresponding to the wells site. 
\begin{center}
\begin{figure}
\includegraphics[width=11cm]{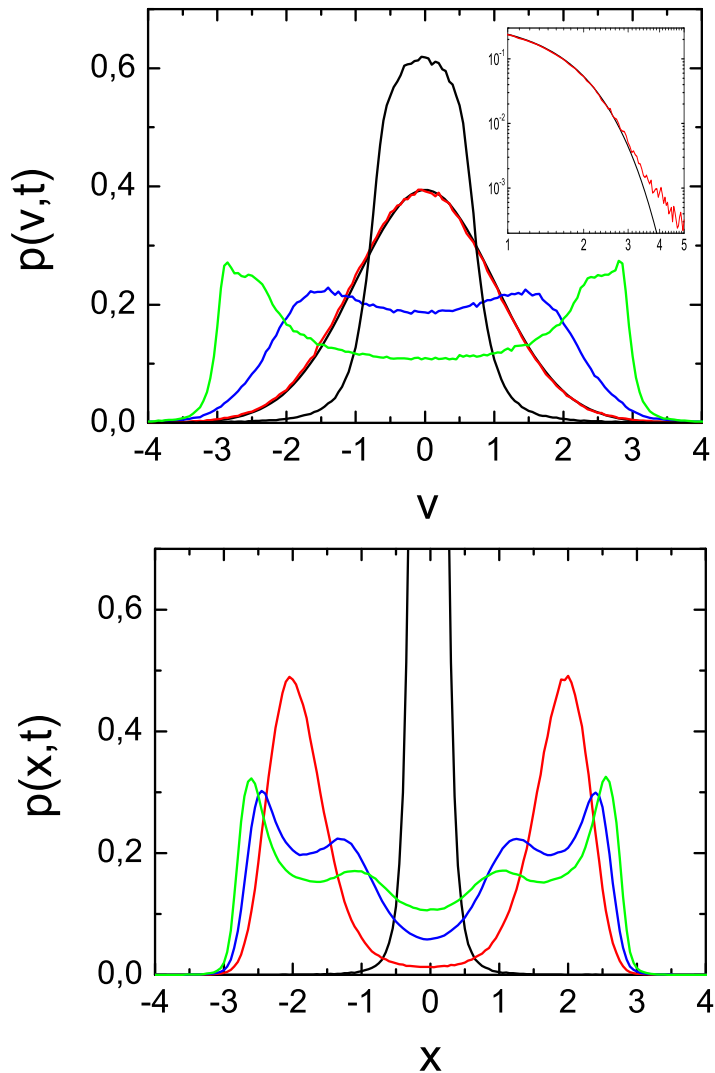}
\caption{The time-evolution of the probability density distributions obtained from Eq.(\ref{gle}) 
for $\theta=2$, $\lambda=1$, $T=1$ and $m=1$. 
Upper part: the velocity distribution for the following times: 
2 (green), 10 (blue), 50 (red) and 1 (black) (from left to right on the left-hand side of the figure). 
The black line which coincides with the case $t=50$ marks the function $\exp(-v^2/2)$. 
Inset: the case $t=50$ in the log-log scale. Lower part: the position distribution 
for the following times: 5 (green), 10 (blue), 40 (red) and 1 (black).}
\end{figure}
\end{center}

In order to make the physical meaning of the results more transparent, we rescale the variables 
to their dimensionless form. The characteristic length is given by the potential size, $L=\sqrt{A/B}$, 
and the characteristic time $\tau_0$ follows from the force balance $mL/\tau_0^2=\Delta V/L$ \cite{lucz} where 
$\Delta V=A^2/4B$ is the barrier height; then we have $\tau_0=2\sqrt{m/A}$. The dimensionless quantities 
are the following: $\bar x=x/L$, $\bar v=\tau_0 v/L$, $\bar t=t/\tau_0$, $\bar\xi=\xi/AL$, $\bar w=mw/AL$, 
$\bar\eta=\eta/AL$, $\bar\lambda=\tau_0\lambda$, $\bar\gamma=\tau_0\gamma$, $\bar c_1=c_1/\sqrt{\tau_0}$, 
$\bar{\cal C}={\cal C}/(AL)^2$ and $\bar D=AL\tau_0D$. Eq.(\ref{yodx}) takes the form 
$\bar \eta=\bar K^{-1}/(1+\theta)|\bar\xi|^{1+\theta}\hbox {sgn}(\bar\xi)$, where $\bar K=K/(AL)^\theta$. 
In the following analysis, we use the dimensionless quantities and drop the bar sign. Eq.(\ref{gled}) 
assumes the form
\begin{eqnarray}
\label{gled1}
\dot v(t)&=&4(x-x^3-w(t)+c_1\sqrt{\lambda}\xi(t))\nonumber\\
\dot w(t)&=&-\lambda w(t)+ AL^2c_1^2{\cal C}(0)\lambda v(t)/T \nonumber\\
\dot x(t)&=&v(t)\nonumber\\
\dot \eta(t)&=&-\gamma(1+\theta)\eta(t)+L(t); 
\end{eqnarray}
moreover, $[m]=$g and $[T]=$g $\hbox{cm}^2/\hbox{sec}^2$. 

\section{Jumping over a potential barrier}

Transport properties of the dynamical systems with the multiplicative noise 
are different than those for the additive noise. The variance, which always is finite for 
the Gaussian processes, may rise with time not only linearly but also slower and faster than that, 
indicating the anomalous diffusion. 
If the stochastic driving obeys the general L\'evy stable statistics, different from 
the Gaussian, the dynamics with the additive noise implies the accelerated diffusion 
since the variance is infinite. This rule may not be valid if one introduces the multiplicative 
factor (\ref{godxi}) to the Langevin equation. In the Stratonovich interpretation, the 
asymptotic form of the probability density distribution is not the same as for the driving noise, 
in contrast to the It\^o interpretation, and depends on $\theta$, 
$p_S(x)\sim|x|^{-1-\alpha-\theta}$. As a consequence, variance may be finite and the system 
subdiffusive \cite{sro09}. 

On the other hand, one can ask how long a particle abides in a given area before it finally 
escapes. This physically important problem has been extensively studied for the Gaussian and Markovian 
processes and, in particular, a transition rate in the double-well potential 
was calculated \cite{mod}. The jumping over a barrier for the case of the L\'evy stable noise 
was analysed in terms of a first passage time distribution in Ref. \cite{dyb}. 
MFPT monotonically rises with $\alpha$ for the symmetric noise if both a reflecting 
and absorbing barrier are assumed. Predictions of the Langevin equation 
with the multiplicative L\'evy stable noise depend, in addition, on 
the parameter $\theta$ and on a specific interpretation of the stochastic 
integral \cite{sro10,sro12b}. MFPT initially falls with $\theta$ and then rises 
which behaviour means -- for the Stratonovich interpretation -- that the dependence 
on the effective barrier width is stronger than on the barrier height. 

We evaluate the time the particle needs to pass from the right to the left well  
assuming the initial condition $p(x,0)=\delta(x-x_m)$. To exclude events 
of reentering the right well, the point $x=-x_m$ is set as an absorbing 
barrier. However, in the presence of jumps particle may skip the barrier position 
without hitting it and then the boundary condition must be assumed as nonlocal 
\cite{dyb}: $p(x<-x_m,t)=0$. Moreover, the particle is reflected from the right flank of the potential. 
The density distribution with the boundary condition, $p(x,t)$, allows us 
to determine time characteristics of the barrier penetration. The survival probability, i.e. 
the probability that particle has not yet reached the absorbing barrier, 
is given by $S(t)=\int_{-x_m}^\infty p(x,t)dx$. 
From this quantity the first passage time density distribution directly follows, 
$f(t)=-dS(t)/dt$, and the averaging over that distribution produces the MFPT:
\begin{equation}
  \label{12}
  T_p=\int_0^\infty tf(t)dt=\int_{-x_m}^\infty dx\int_0^\infty p(x,t)dt.
\end{equation} 

In the following, we demonstrate how passage statistics depends on the model parameters. 
Fig.2 presents the survival probability for a fixed temperature and some values of 
$\lambda$, $\theta$ and $m$. The shape is exponential for all the cases but the rate strongly 
depends on the parameters, rising with $\lambda$ and diminishing with $m$. 
Dependence on $\theta$ for small $T$, like the case shown in the figure, 
is rather complicated. The slope rises with $\theta$ when $\theta$ is large but 
it becomes very small for small values of this parameter. 
The exponential shape of $S(t)$ ensures that MFPT is always finite. 
\begin{center}
\begin{figure}
\includegraphics[width=11cm]{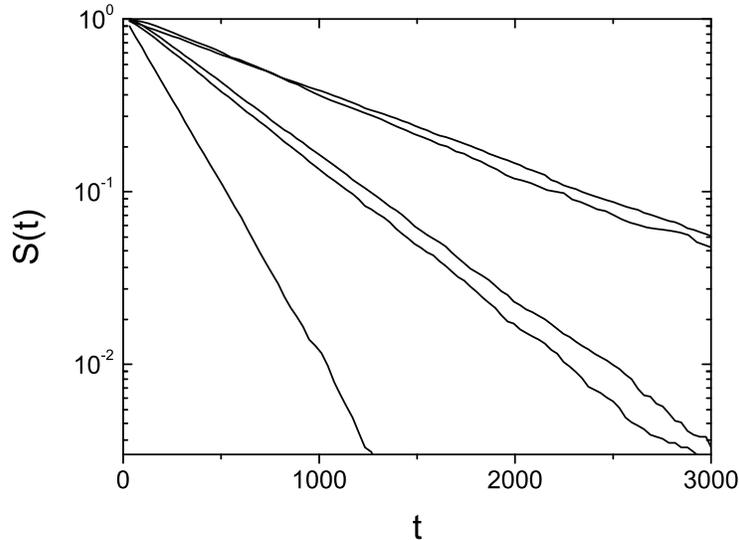}
\caption{The survival probability for $T=1$. The following cases are presented: 
1. $\lambda=0.632$, $\theta=2$ and $m=0.1$; 2. $\lambda=0.316$, $\theta=2$ and $m=0.1$; 
3. $\lambda=0.316$, $\theta=4$ and $m=0.1$, 4. $\lambda=0.316$, $\theta=2$ and $m=0.5$; 
5. $\lambda=0.316$, $\theta=1.2$ and $m=0.1$ (from left to right).}
\end{figure}
\end{center}

One can expect that MFPT diminishes with the temperature which effect is presented in Fig.3 
for some values of $\lambda$ and a fixed $\theta$ ($\gamma$ is different 
for each value of $\lambda$). MFPT is very large for small values of $T$ and then rapidly falls. 
Finally it stabilises when the average kinetic energy is sufficiently large to make possible 
an easy passage above the barrier. The case of the largest memory time,  
$1/\lambda$, is characterised by a large MFPT and, on the other hand, 
MFPT becomes small in the white-noise limit. This result may be attributed to 
a large noise intensity for a large $\lambda$. 
The function $T_p(\theta)$ behaves similarly to $T_p(\lambda)$ -- 
the largest $T_p$ is observed for small $\theta$ -- and that dependence is presented 
in Fig.4. MFPT decreases with $\theta$ for a given $T$ but only if $\theta$ 
is relatively small; then it saturates. This behaviour reflects the cosine dependence 
of the noise intensity on $\theta$: small values of $\theta$ correspond to small 
noise intensity which approaches zero in the limit $\theta\to1$. 
The above observation does not strictly hold in the region of small $T$ 
since there large values of $\theta$ may lead to relatively large $T_p$. 
This effect is visible also in the next figure. 
\begin{center}
\begin{figure}
\includegraphics[width=11cm]{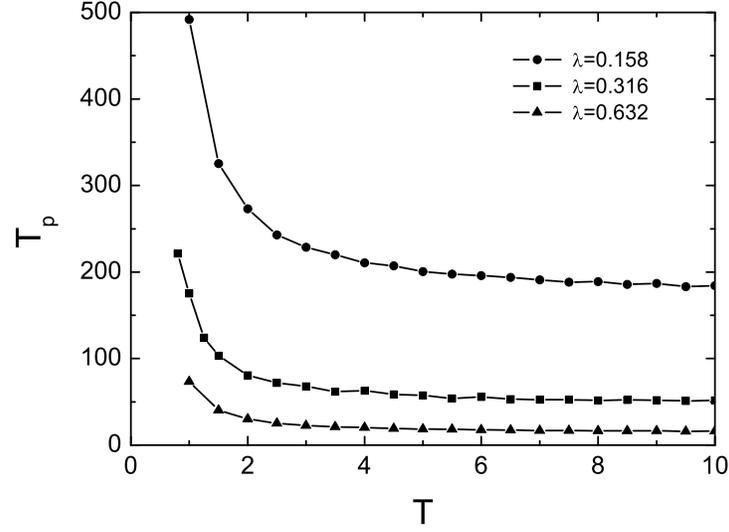}
\caption{MFPT as a function of the temperature for $\theta=2$ and $m=0.1$.}
\end{figure}
\end{center}
\begin{center}
\begin{figure}
\includegraphics[width=11cm]{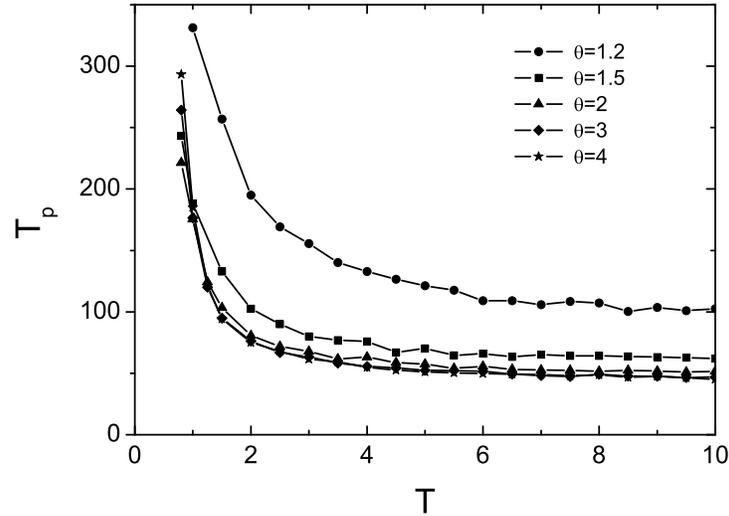}
\caption{MFPT as a function of the temperature for $\lambda=0.316$ and $m=0.1$.}
\end{figure}
\end{center}

Fig.5 comprises the dependences on all parameters. Since the sensitivity on the potential structure 
is particularly pronounced at the small temperature, a relatively low value, $T=1$, 
was chosen. As one may expect, MFPT monotonically increases with the mass and is large 
for large memory, in agreement with Fig.3. The dependence on $\theta$ for a given $m$ 
is more complicated. MFPT is largest for a small $\theta$ but then this trend turns to 
the opposite: the minimum of MFPT corresponds to $\theta=2$ and then its value rises again. 
Though most of the curves falls monotonically with $\lambda$ and saturates for its large value, 
the case $\theta=4$ exhibits a clear minimum at $\lambda=1.2$. This non-trivial behaviour may reflect 
a different dependence of the noise intensity and the damping on both parameters. 
\begin{center}
\begin{figure}
\includegraphics[width=11cm]{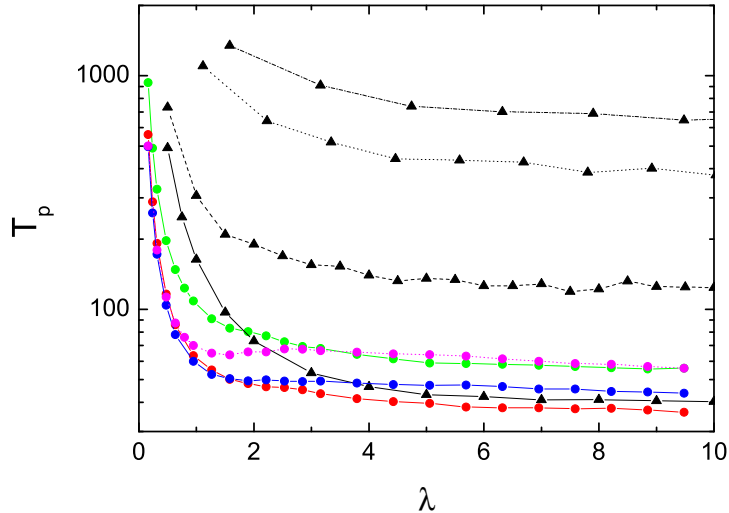}
\caption{MFPT as a function of $\lambda$ for $T=1$ and different sets of parameters: 1. $\theta=2$ 
with $m=10, 5, 1, 0.1$ (triangles, from top to bottom), 2. $m=0.1$ with 
$\theta=4$ (magenta), 1.2 (green), 3 (blue), 1.5 (red) 
(points, from top to bottom on the right hand side).}
\end{figure}
\end{center}

\section{Stochastic resonance}

The possibility that a weak signal may be strengthened by a random perturbation 
makes the stochastic resonance problem interesting from the point of 
view of applications. That phenomenon \cite{benz} consists in matching the noise-induced 
rate of jumping between the potential wells and the frequency of an external periodic 
force. It was extensively discussed for the Gaussian noise \cite{gamm1} and generalised 
to the L\'evy flights \cite{kos,dybr,dybr1}. SR is observed also 
in systems characterised by a multiplicative noise with long jumps \cite{sro12b}. 

Studies of SR are not restricted to the Markovian processes. The non-Markovian features of 
the dynamics in the context of SR were observed in the framework of a two-state bistable 
model \cite{goy} which was a generalisation of the well-known McNamara and Wiesenfeld 
theory \cite{mcn}. The residence time distribution appears non-exponential for the 
non-Markovian case: its shape assumes a stretched-exponential, or even a power-law, form. 
The SR strength becomes strongly suppressed by the non-Markovian effects \cite{goy}. 
This strength used to be measured by a spectral amplification or a signal-to-noise ratio as 
a function of the temperature. SR was observed in the 
non-Markovian dynamics for an external, instantaneous noise \cite{ful} and 
was studied also for GLE \cite{jun1,hae,nei}. 
Properties of SR are strongly influenced by the memory strength (the noise intensity) and time. 
SR is suppressed by the memory in the overdamped limit but if the coloured noise is induced 
by inertia, which leads to the Kramers equation, one obtains an enhancement of SR \cite{hae}. 
Similar results were obtained in Ref. \cite{nei} where the kernel 
comprised an instantaneous friction term and the exponential function. 
It has been demonstrated that, if the memory time is large, 
the power amplification rises with both the memory strength and time. 

We shall demonstrate that GLE involving jumps also predicts emergence of SR. 
We consider Eq.(\ref{gle}) where the deterministic force $F(x,t)$ consists of the bistable 
part (\ref{pot}) and a time-dependent oscillatory force: 
\begin{equation}
\label{pott} 
F(x,t)=-\partial V(x)/\partial x + A_0\cos(\omega_0t), 
\end{equation} 
where the amplitude $A_0$ and the frequency $\omega_0$ are constant and dimensionless. The density 
distribution $p(x,t)$ follows from GLE and can be determined by solving the equations (\ref{gled}). 
SR emerges if the GLE solution, $x(t)$, is correlated with the periodic stimulation 
in such a way that the asymptotic amplitude of the output signal, $\bar x$, exhibits a maximum 
when plotted as a function of the noise intensity. This quantity can be expressed in terms 
of the first Fourier coefficient of the correlation function in the form \cite{jun} 
\begin{equation}
\label{wsp}
\bar x=Re\frac{\omega_0}{\pi}\int_{-\infty}^\infty dx\int_0^T xp_{as}(x,t)\exp(i\omega_0t)dt,
\end{equation} 
where $T=2\pi/\omega_0$. The distribution $p_{as}(x,t)$ is a long-time limit of $p(x,t)$ and 
corresponds to the periodic asymptotic solution of GLE. 
Then, $\bar x$ allows us to define the spectral amplification, 
$\eta=4(\bar x/A_0)^2$, which represents a ratio of the integrated 
power stored in spikes of the power spectrum to the total power carried by 
the oscillatory force \cite{gamm1}. 
\begin{center}
\begin{figure}
\includegraphics[width=11cm]{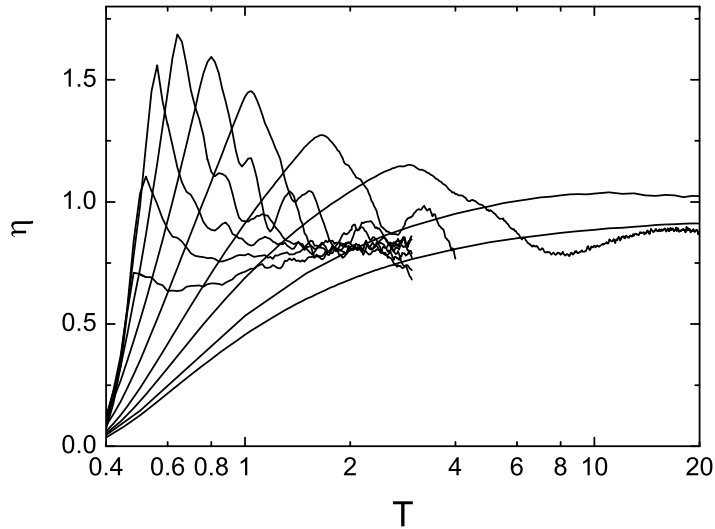}
\caption{The spectral amplification for $m=0.1$, $\theta=2$, $\lambda=0.316$  
and the following values of $A_0$: 1.1, 1.15, 1.225, 1.35, 1.5, 1.65, 1.85, 2, 2.15 and 2.25 (from left to right).}
\end{figure}
\end{center}

Since response of the system is nonlinear in respect to the system parameters, 
the spectral amplification depends on the forcing amplitude $A_0$; $\eta$ decreases 
with $A_0$ and when it becomes large, the resonance vanishes. On the other hand, 
for a small $A_0$, $\eta$ approaches a value which follows from the linear response 
theory. However, if the driving frequency $\omega_0$ exceeds the Kramers rate, 
$\eta(A_0)$ exhibits a maximum \cite{gamm1}. Results for our system are presented in Fig.6. 
In all the calculations we assume $\omega_0=5\tau_0$ which quantity is relatively large to make 
the averaging interval small and then to reduce the computation time. Since $\eta$ decreases 
with $\omega_0$ \cite{gamm1}, one can expect a moderate amplification. The averaging was 
started after $t=50/\tau_0$ to reach the convergence of $p(x,t)$ to $p_{as}(x,t)$ and to ensure 
that transients are not present. Results shown in Fig.6 indicate that the amplification 
decreases with $A_0$ for large values of this parameter, the peak shifts toward large $T$ 
and finally, for $A_0>2.2$, the resonant behaviour is no longer observed. The function 
$\eta(A_0)$ reaches a maximum at $A_0=1.35$, it decreases when $A_0$ gets smaller and 
the resonance vanishes at about $A_0=1.1$. From now on, we assume $A_0=1.5$. 

In the following, we demonstrate how the presence and properties of SR, represented 
by the function $\eta(T)$, depend on the system parameters. 
Fig.7 presents $\eta(T)$ for some values of $\lambda$. 
A peak, indicating a presence of SR, is observed in all the cases and 
the figure demonstrates that the power amplification is a decreasing function of $\lambda$ 
in agreement with the Gaussian case in the regime of the large memory time \cite{hae,nei}. 
When $\lambda$ becomes small, the curves acquire an oscillatory structure and the peak 
finally disintegrates. This phenomenon is observed in the figure for a very large 
memory time, $\lambda=0.158$. On the other hand, the oscillations vanish in the white-noise 
limit (large $\lambda$). The position of the peak shifts to the right with increasing $\lambda$. 
\begin{center}
\begin{figure}
\includegraphics[width=11cm]{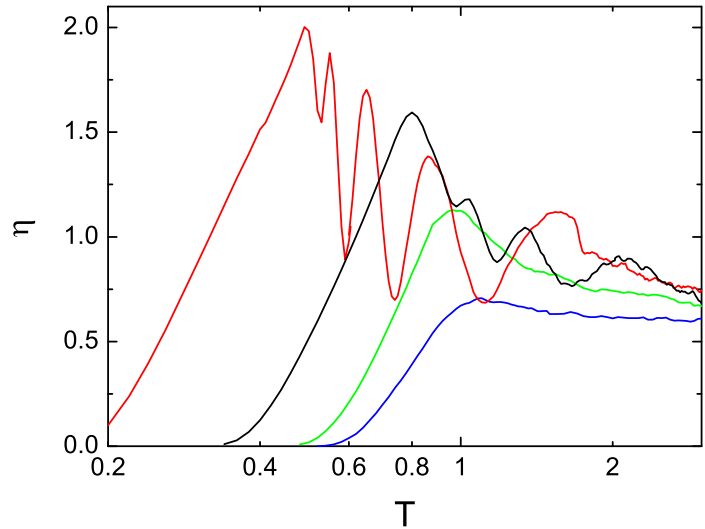}
\caption{The spectral amplification for $m=0.1$, $\theta=2$ 
and the following values of $\lambda$: 0.158, 0.316, 0.474 and 0.632 (from left to right).}
\end{figure}
\end{center}
\begin{center}
\begin{figure}
\includegraphics[width=11cm]{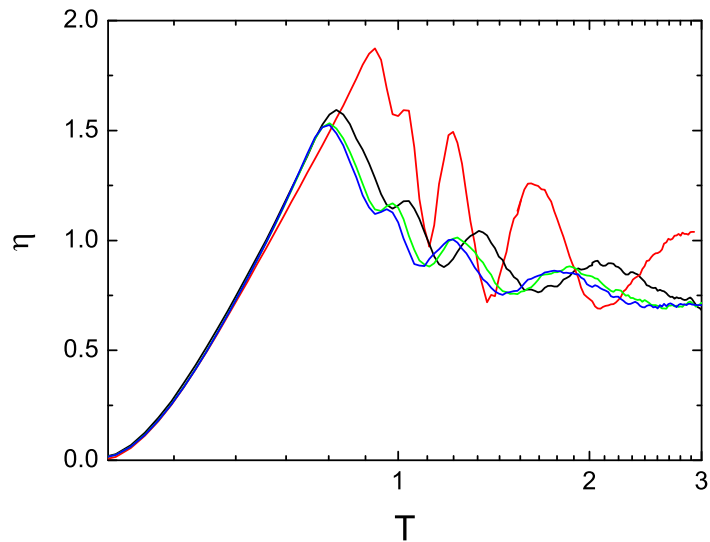}
\caption{The spectral amplification for $m=0.1$, $\lambda=0.316$ 
and the following values of $\theta$: 1.2, 2, 3 and 4 (from top to bottom).}
\end{figure}
\end{center}

The parameter $\theta$ is responsible for the slope of the noise distribution, namely a large 
$\theta$ makes it steeper and reduces the variance of $\xi(t)$. Moreover, it influences 
the memory time $\lambda$. The dependence of the power amplification on $\theta$ 
for a fixed $\lambda$, which means that $\gamma$ depends on $\theta$, is presented in Fig.8. 
The results are sensitive on $\theta$ when it approaches the value $\theta=1$, 
corresponding to the vanishing noise intensity; then $\eta$ 
is largest and oscillations are strong. For larger $\theta$, dependence of $\eta(T)$ on 
$\theta$ is very weak. 
\begin{center}
\begin{figure}
\includegraphics[width=11cm]{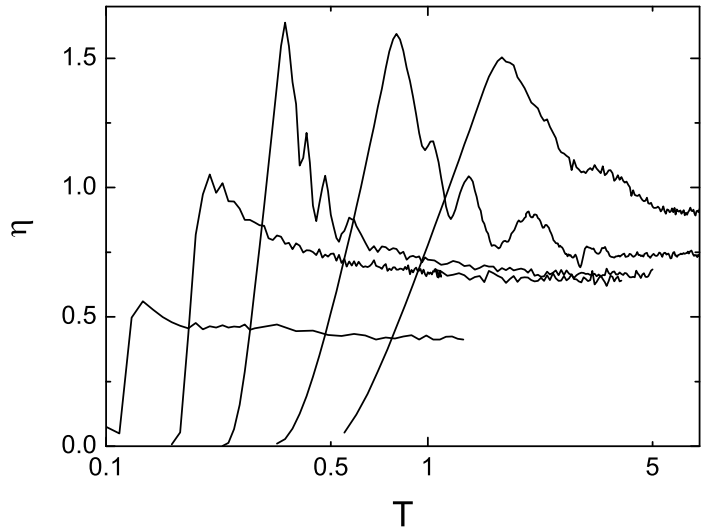}
\caption{The spectral amplification for $\lambda=0.316$, $\theta=2$ 
and the following values of $m$: 0.09, 0.1, 0.12, 0.15 and 0.2 (from right to left).}
\end{figure}
\end{center}

Finally, we consider the dependence of the spectral amplification of the inertia and those 
results are presented in Fig.9. Large values of $m$ correspond to a small intensity 
of SR and smooth peaks with a sharp cut-off at the left-hand side. 
For simple systems, the slopes of the peaks on the right-hand side 
are determined by the potential: they fall to zero if it is asymptotically strong and 
may rise for soft potentials \cite{hei}. In the present case, 
the slopes stabilise at a constant value. For a smaller mass, the peaks 
are wider and their height rises when $m$ decreases but finally it 
saturates. The SR position shifts to the left with the increasing $m$. 

\section{Summary and conclusions}

The characteristic features of the generalised Ornstein-Uhlenbeck process $\xi(t)$, 
including the multiplicative Cauchy noise, are long tails of the distribution, 
a finite variance and a finite correlation time. 
Those properties are typical for the complex systems and 
the multiplicative noise reflects a dependence of the random force on the process value, 
which may result, in particular, from a nonhomogeneity of the space.
The process $\xi(t)$ is defined by two parameters: the damping $\gamma$ and the multiplicative 
noise parameter $\theta$. Both of them determine the relaxation rate $\lambda$, as well as 
the variance which decreases with $\theta$ and $\gamma$. 
If $\xi(t)$ represents an internal effective noise in a dynamical system, 
its finite correlation time requires a retarded friction to satisfy 
the fluctuation-dissipation theorem. We discussed the dynamics governed 
by GLE in which, beside $\xi(t)$, the double-well potential was included. 
The solutions of GLE produce the velocity density distributions which converge to 
a distribution resembling the Gaussian. However, it possesses the power-law tails, 
similarly to the case without any potential. In the latter case, when the system 
is linear, this apparent violation of the central limit theorem is 
an effect of large higher moments. 

The transport properties, quantified in terms of MFPT, were discussed for all model parameters. 
MFPT declines with $T$ and $\lambda$ whereas it rises with $m$. The dependence on $\theta$ is 
not monotonic and complicated for large $\theta$ being different for the low and high temperature. 
The survival probability always reveals the exponential form. 

When the deterministic potential is supplemented by a time-dependent oscillatory force, 
the stochastic resonance emerges. It was studied by evaluating the spectral amplification $\eta$ 
as a function of the temperature. Both the intensity of SR and its position strongly depend on the memory 
parameter $\lambda$: the smaller $\lambda$ the stronger amplification and the peak positioned at 
a lower temperature. The variability of the above quantities with $\theta$, in turn, 
is less pronounced and actually restricted 
to a region of small values of this parameter where the amplification is large. 
A striking feature of the curves corresponding to small values of either $\lambda$ or $\theta$ 
is their oscillatory structure which is observed neither for GLE in the Gaussian 
case \cite{nei} nor for the Langevin equation with the multiplicative noise with jumps \cite{sro12b}. 
This structure emerges when the noise intensity is small and memory long, 
i.e. when trajectories are trapped in the well for a long time (large MFPT). 
The dependence of the SR properties on the inertia is simple and 
the intensity of SR rises when the mass becomes smaller but finally $\eta$ stabilises 
at a relatively large value.

\end{document}